# Deciphering Majorana Zero Modes in Topological Superconductor $FeTe_{0.55}Se_{0.45}$ with Machine-Learning-Assisted Spectral Deconvolution

*Jewook Park[1,†*], Hoyeon Jeon[1,†], Dongwon Shin[2], Guannan Zhang[3], Michael A McGuire[2], Brian C. Sales[2] and An-Ping Li[1*]*

*Center for Nanophase Materials Sciences, Oak Ridge National Laboratory, Oak Ridge, TN 37831, USA*

*[2]Materials Sciences and Technology Division, Oak Ridge National Laboratory, Oak Ridge, TN 37831, USA*

*[3]Computer Science and Mathematics Division, Oak Ridge National Laboratory, Oak Ridge, TN 37831, USA*

## Abstract

Unambiguous identification of Majorana zero modes (MZMs) in topological superconductors (TSCs) remains a challenge due to complex in-gap states that can also produce zero-bias conductance peaks (ZBPs). Here we demonstrate a data-driven workflow that integrates pixel-wise spectral deconvolution with machine-learning (ML) to analyze tunneling spectroscopy from $FeTe_{0.55}Se_{0.45}$, an intrinsic TSC. Based on the local density of states (LDOS) spectra acquired with a millikelvin scanning tunneling microscope under magnetic fields, each spectrum was decomposed into multiple Lorentzian peaks. The extracted peak parameters were assembled into a structured feature set and subsequently embedded and clustered with unsupervised ML algorithms. ML-based clustering identified distinct classes of LDOS spectra, separating superconductor vortices exhibiting ZBPs consistent with established characteristics of MZMs from vortices displaying ZBP-mimicking features of trivial origin. Furthermore, spatially resolved ZBP distributions differentiate isotropic vortex cores with well-defined ZBPs from vortices that exhibit locally distorted ZBPs. By comparing the ZBP distributions to defect locations measured without magnetic field, we found a correlation between local heterogeneity and the ZBP formation, necessitating the systematic, data-driven analysis to disentangle genuine MZM signatures in TSC. This objective and reproducible workflow advances reliable MZM detection in TSCs, providing a foundation for MZM manipulation towards quantum computation.

[†] These authors contributed equally.

* Corresponding authors: parkj1@ornl.gov, apli@ornl.gov

# Introduction

Majorana zero modes (MZMs) are zero-energy quasiparticle excitations that emerge at boundaries, defects, or vortex cores of topological superconductors (TSC). The non-Abelian statistics of MZMs offer a promising route towards fault-tolerant quantum computation[1]. $FeTe_{0.55}Se_{0.45}$ (FTS) has been proposed as an intrinsic TSC[2] and tunneling spectroscopy has revealed zero-bias peak (ZBP) at the vortex core[3], suggesting the presence of MZMs. FTS has a relatively large superconducting gap ($\Delta_{sc} \approx 1.8$ meV)[4] and small Fermi energy ($E_f \approx 4.4$ meV)[2], giving level spacing of the topologically trivial Caroli-de Gennes-Matricon (CdGM) states ($E_\mu = \mu\Delta_{sc}^2/E_f$) on the order of hundreds of μeV[3, 5]. In principle, ZBPs can be resolved from CdGM states at sufficiently low temperature. Indeed, ultra-low temperature scanning tunneling microscopy and spectroscopy (STM/S) has been employed as an essential tool in the search for MZMs, which have revealed sharp, non-splitting ZBPs in a subset of vortices, consistent with the expected signature of MZMs[3, 6, 7]. The ZBPs as an MZM signature has been further strengthened by ruling out several non-topological mechanisms, reflectionless tunneling, Kondo effect, Josephson supercurrent, and weak antilocalization based on a sharp, spatially non-dispersive, and field-robust ZBP in a vortex core[3]. Additional evidence has been reported consistent with MZMs, such as a nearly quantized conductance plateau[8] and spatial distribution of MZM hosting vortices[7, 9]. However, alternative spectroscopic features have been reported as well, complicating the initial interpretation of ZBPs in FTS[6, 10, 11]. For example, excess Fe atoms host robust zero-energy bound states with trivial origins[12, 13], domain boundaries can display finite in-gap states[14, 15], in a certain case discrete CdGM levels are observed in vortex cores but without ZBPs[5], plateau-like features can also arise in trivial Yu–Shiba–Rusinov (YSR) states[6, 10], and collective YSR impurity states can form an amorphous in-gap spectral background[16]. Moreover, impurity in TSC can shift CdGM levels arbitrarily close to zero energy, leading to apparent ZBPs that mimic MZM signature[17]. Hence, MZM detection in FTS remains challenging even with high energy resolution at low temperature STM/S, demanding comprehensive examination of the spectral and spatial evolution of near-zero-energy states across the relevant spatial range, rather than relying on isolated zero-bias features[11].

Addressing this challenge requires not only achieving the high energy resolution to precisely isolate ZBPs from non-topological near-zero-energy states (further explanation in Supplementary Information 1) but also characterizing the spatial evolution of the ZBPs within the vortex to compare with theoretical calculations of MZMs. Previous STM/S studies examined ZBPs by using line[3] or point[7, 18] spectroscopy, which sample only a limited subset of the available data. Analyzing the complete datasets of spatial and spectral distributions of local density of states (LDOS, $\rho(E, \boldsymbol{r})$) would enable tracking their spatial evolution across entire vortex cores. This pixel-by-pixel data handling calls for machine-learning (ML)-assisted approaches. ML helps objectively identify ZBPs as MZM candidates and separate them from other in-gap states with trivial origins. Although ML-assisted efforts have recently been made to improve the identification of ZBP signatures associated with MZMs in hybrid nanowires[19-21], such approaches have not yet been established for experimental STM/S datasets of ZBPs in intrinsic topological superconductors.

Here, we report a data-driven workflow that combines high spatial and energy resolution STM/S data acquisition, spectral deconvolution with multiple peak fitting, and unsupervised ML to isolate ZBP in FTS. Using a millikelvin STM operating at 40 mK, we acquired an array of dI/dV spectra over a regular grid to construct spatially resolved, bias-dependent LDOS maps, hereafter grid LDOS data. The grid LDOS data are statistically analyzed to extract spatially resolved spectral features using an unsupervised ML algorithm, which identifies ZBP-related clusters and provides an objective basis to reconstruct grid LDOS data highlighting the spatial distribution of ZBPs. The resulting ZBPs are evaluated against experimental criteria for MZM signatures to clarify their topological origin and exclude trivial bound states such as disorder-shifted CdGM levels, YSR states, and subsurface contributions. Importantly, our goal is not to provide a complete classification of every local spectroscopic feature, but to reliably extract ZBP spectral features from grid LDOS data and identify MZM-candidate vortices through unsupervised clustering. The ML-assisted workflow strengthens statistical inference across large spectroscopic datasets, and it offers a scalable framework adaptable to complementary measurements, such as the nonlocal transport[22] and spin-polarized STM[23], to strengthen MZM identification (Supplementary Information 2).

## Results and discussion

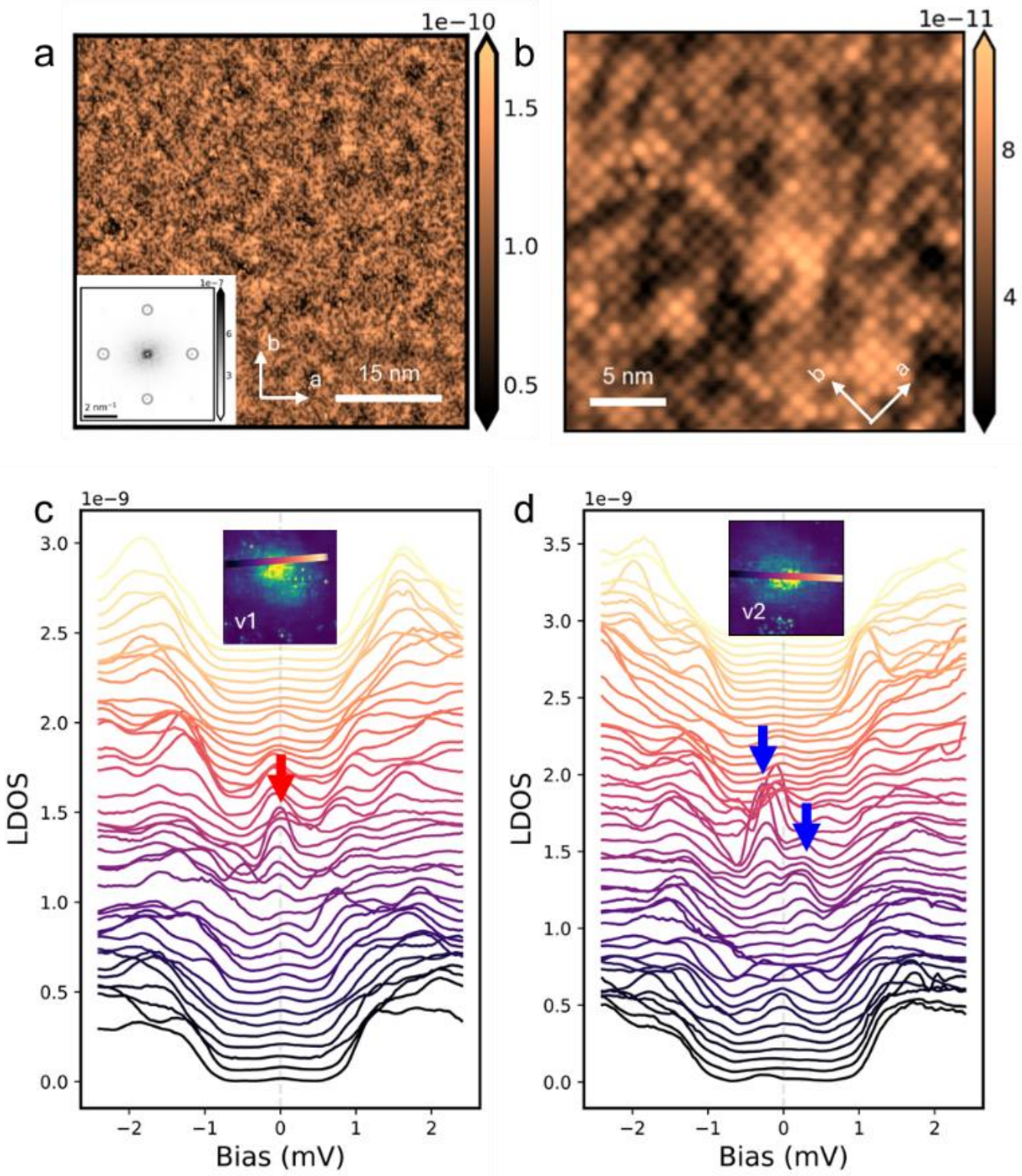

**Figure 1**. Scanning tunneling microscopy (STM) topography and spectroscopy across the vortex core of $FeTe_{0.55}Se_{0.45}$ (FTS). (a) STM image (64 × 64 $nm^2$, $V_b = -4$ mV, $I_t = 400$ pA) acquired at 40 mK with corresponding 2D fast Fourier transform (FFT) image (inset). Gray circles in the FFT mark the Bragg points from the topmost Se/Te lattice. (b) High-resolution STM image (10 × 10 $nm^2$, $V_b = 20$ mV, $I_t = 200$ pA) from the same sample. The local height corrugation of the top chalcogen atoms indicates a chemical contrast between Se and Te. (c, d) Stacked dI/dV spectra acquired along a line across (c) vortex1 (v1) and (d) vortex2 (v2). The insets show zero-bias conductance (ZBC) map with horizontal line indicating the line-profile locations. The dI/dV spectrum colors indicate spatial positions along the line with matching dI/dV spectra color. A pronounced zero-bias peak (ZBP) is observed at the core of v1 (red arrow), whereas near-zero-energy features are observed off-center within the core of v2 (blue arrows).

The FTS single crystal[24] was cleaved at low temperature (~82 K) under ultrahigh vacuum (~$1\times10^{-10}$ Torr) and immediately transferred to a precooled STM head. All STM/S measurements were performed with a dilution-refrigerator STM equipped with high field magnets, using normal metal tips. We achieved high energy resolution (Supplementary Information 1) to distinguish the expected CdGM levels and ZBPs[9] at a base temperature of 40 mK. We focused on atomically clean surfaces without Fe adatoms[12] or domain boundaries[14] to avoid contributions of local heterogeneities, and Figure 1a displays a large scale STM topography with its fast Fourier transform, confirming the absence of excess Fe adatoms and domain boundaries. Figure 1b shows atomic corrugations of Se/Te lattice[4, 25].

Despite these controlled conditions, the tunneling spectrum still exhibits intricate in-gap states from disorder, such as subsurface defects. Figures 1c, d illustrate stacked dI/dV spectra across simultaneously acquired two vortices at $B_{\perp} = 2$ T. Both spectra exhibit clear superconducting gaps, yet their in-gap states are different. Vortex 1 (Fig. 1c) hosts a pronounced ZBP at the core (red arrow), whereas vortex 2 (Fig. 1d) exhibits near-zero-energy features (blue arrows). Variations in subgap conductance can arise from subsurface defects[13, 15], which produce near-zero-energy states even without magnetic field. Also, defects can generate local electrostatic inhomogeneities to shift CdGM states toward near-zero energy[17], thereby giving rise to ZBP-mimicking features.

Accordingly, deciphering an MZM-hosting vortex goes beyond the observation of a non-splitting ZBP at the vortex core. While enhanced ZBC and ZBPs at the vortex core remain essential indicators, analyzing the spatial and spectral distributions of the LDOS near zero bias is required to further distinguish the complex overlap of in-gap states. ZBP as an MZM signature is expected to be strongly localized at the vortex core and remain non-splitting, while the CdGM states are expected to exhibit energy-dependent spatial distributions. To spatially resolve in-gap states, spectroscopic features are first deconvoluted, and their spatial distributions are subsequently examined as a function of magnetic field. This requires deconvolving multiple peaks in LDOS and analyzing their spatial distributions over full grid LDOS data, rather than a limited number of representative spectra. In general, manual inspection of overlapping spectra features and investigation of their spatial correlations is impractical across thousands of local spectra, particularly in the presence of vortex-to-vortex variations and heterogeneous subsurface disorder. This motivates a data-driven workflow that systematically organizes and classifies the deconvoluted peak components across the entire grid LDOS data sets to disentangle MZM-related ZBPs from trivial vortex core states.

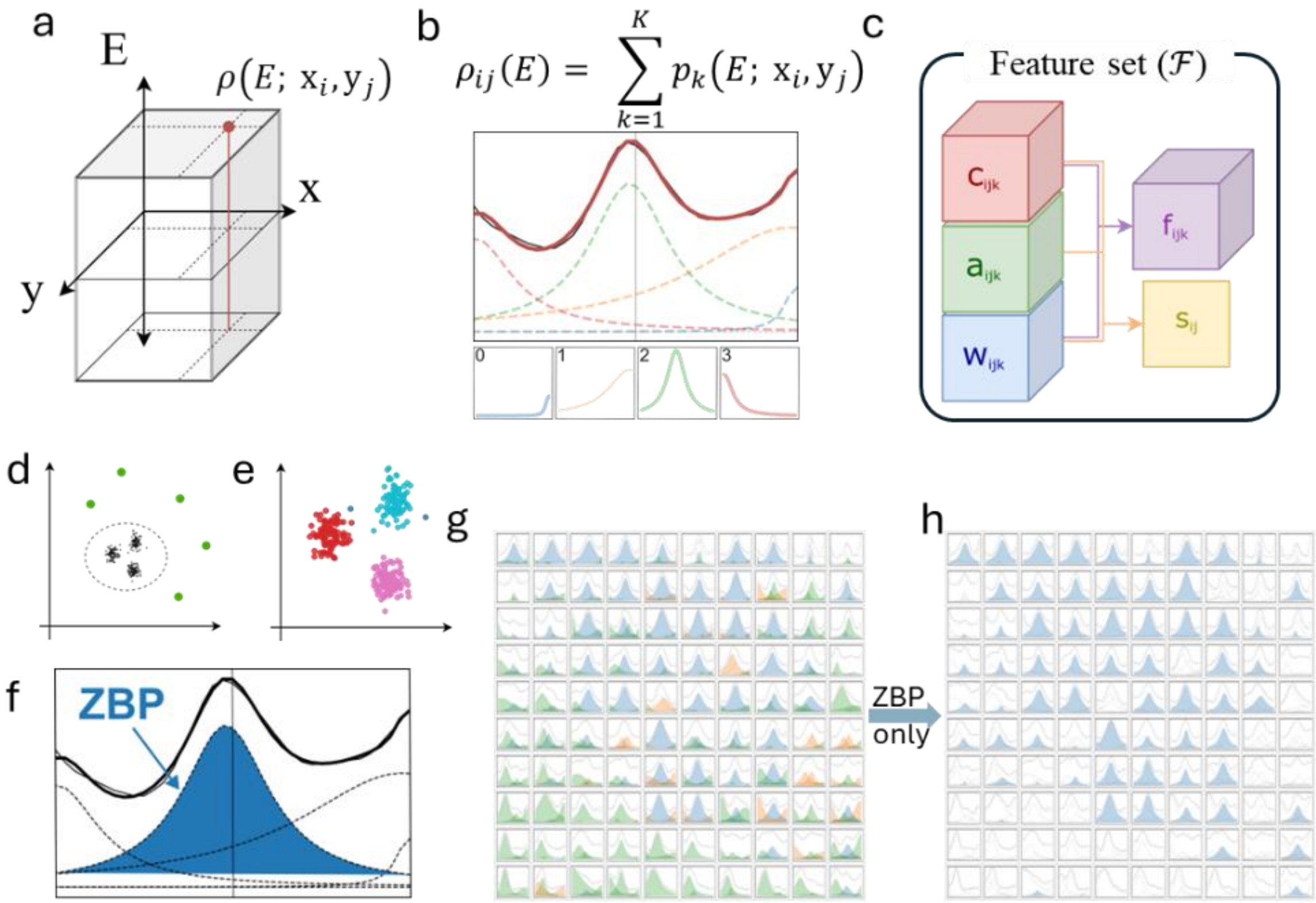

**Figure 2.** Machine-learning (ML) workflow for ZBP detection from grid local density of states (LDOS) data. (a) Illustration of STM/S measuring grid LDOS data $\rho(E, \boldsymbol{r})$ as a function of energy and spatial coordinates. (b) Local spectrum $\rho_{ij}(E)$ deconvoluted into multiple Lorentzian peaks. (c) The extracted peaks parameters (center $c_{ijk}$, amplitude $a_{ijk}$, and width $w_{ijk}$) stacked into a multi-dimensional feature set $\mathcal{F}$, and augmented with additional features that capture zero-bias proximity and spectral symmetry (Supplementary Information 4). (d, e) Unsupervised ML steps including (d) outlier removal and (e) clustering process. (f) A ZBP-consistent cluster identified by quantitative cluster statistics, and the corresponding peaks are assigned as ZBP components. (g) Cropped 10×10 array of LDOS spectra in spatial grid layout and their deconvoluted peaks. At every grid point $\mathbf{r}_{ij}$, a miniature spectrum $\rho_{ij}(E)$ with its Lorentzian fit $p_k(E;\ \boldsymbol{r}_{ij})$ are displayed. The shading colors of each $p_k(E;\ \boldsymbol{r}_{ij})$ represents its ML-assigned cluster labels ($C_0$: blue, $C_1$: orange, and $C_2$: green, Supplementary Information 5). (h) ML-assisted filtering retaining only the ZBP peaks, demonstrate the isolation of ZBP component from complex LDOS spectra.

Figure 2 illustrates our unsupervised ML-assisted workflow that analyzes grid LDOS data through spectral deconvolution to isolate ZBPs, filters out trivial in-gap states, highlights non-splitting ZBPs at vortex cores, and enables us to identify putative MZM signatures. The grid LDOS data $\rho(E, \boldsymbol{r})$ are sampled over spatial coordinates $\mathrm{r}_{ij} = \left(\mathrm{x}_i, \mathrm{y}_j\right)$ and bias voltage E, forming a three-dimensional (3D) dataset (Fig. 2a). Each local dI/dV spectrum $\rho_{ij}(E)$ is deconvoluted into a sum of K Lorentzian peaks, $\rho_{ij}(E) = \sum_{k=1}^{K} p_k\left(E;\ \boldsymbol{r}_{ij}\right)$, where $p_k\left(E;\ \boldsymbol{r}_{ij}\right) = Lorentz\left(E; c_{ijk}, a_{ijk}, w_{ijk}\right)$ and $c_{ijk}$, $a_{ijk}$, $w_{ij}$ denote the peak center, amplitude, and width, respectively. Lorentzian line shapes reflect quasiparticle lifetime broadening in tunneling spectroscopy (Supplementary Information 3). All extracted peak parameters are stacked into a multi-dimensional feature set $\mathcal{F} \in \mathbb{R}^{N \times M \times K \times D}$, where N and M correspond to the spatial grid dimensions along x and $y$, K is the number of peaks per spectrum, and D is the parameter dimension. In addition to the fitted peak parameters, we augment the feature space with heuristically defined, physically interpretable descriptors, such as zero-bias proximity and spectral symmetry, chosen to enhance separability in the embedding space (Supplementary Information 4). For the subsequent unsupervised ML analysis, clustering is performed on the feature vectors of individual peaks. Spatial coordinates are deliberately excluded from the embedded feature space, so that the clustering reflects intrinsic spectral characteristics.

To ensure robust and objective classification of spectral features with unsupervised ML, statistical outliers are excluded (Supplementary Information 4). Principal component analysis (PCA) and k-nearest neighbor (kNN) distance scoring are used to identify outliers in PCA space (Fig. 2d). The cleaned features are embedded using uniform manifold approximation and projection (UMAP)[26], and clustered using hierarchical density-based spatial clustering of applications with noise (HDBSCAN)[27] (Fig. 2e). A cluster consistent with ZBPs is identified through quantitative cluster statistics, and the corresponding deconvoluted peaks are assigned as ZBP components in each LDOS spectrum (Fig. 2f). Figures 2g, h present a cropped 10×10-pixel portion of the grid LDOS data, displayed as an array of local spectra to illustrate how the ML workflow disentangles complex spectral components. Each small panel corresponds to one spatial pixel and shows LDOS spectrum $\rho_{ij}(E)$ together with deconvoluted peaks $p_k\left(E, \boldsymbol{r}_{ij}\right)$. All decomposed peaks from the 10×10-pixel region are displayed in Fig. 2g, whereas Fig. 2h shows only the ZBP cluster (blue shading). This demonstrates that the ML workflow can effectively isolate and map the ZBP contribution from complex LDOS spectra (Supplementary Information 5).

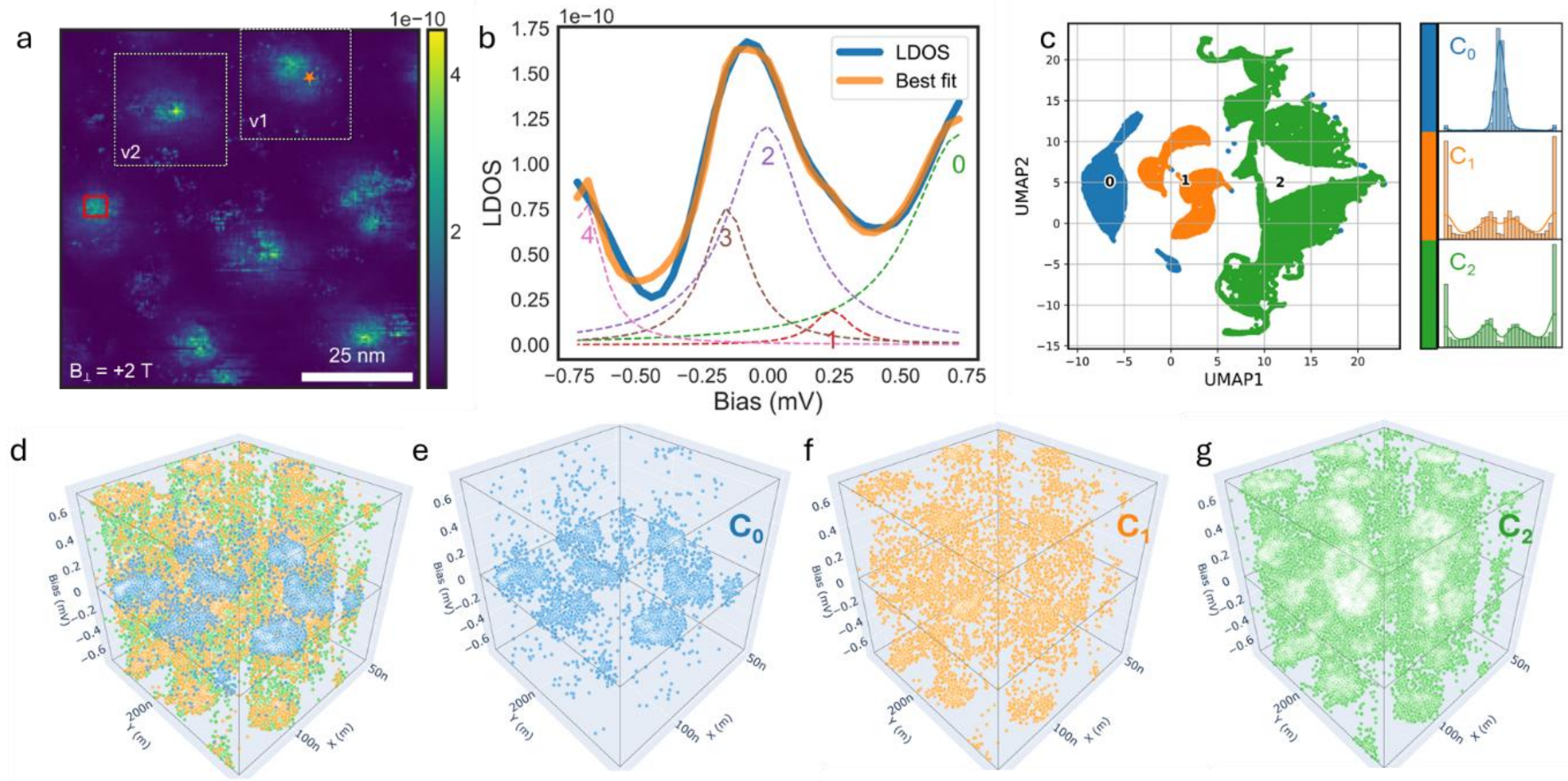

**Figure 3**. ML-assisted identification of ZBPs in FTS. (a) ZBC map ($\rho(E = 0, \boldsymbol{r})$, $80 \times 80$ nm$^2$) extracted from grid LDOS data of FTS under $B_\perp = 2$ T. White boxes mark regions v1 and v2 in Figs. 1c and 1d, respectively. Red box marks the region with the cropped array of local LDOS spectra shown in Fig. 2g. (b) Example LDOS spectrum at orange star in (a) with multiple Lorentzian fitting results within superconducting gap. Black and red solid lines denote the measured LDOS spectrum and best fit result, respectively, dashed lines represent the deconvoluted Lorentzian peaks. (c) Uniform manifold approximation and projection (UMAP) embedding of all deconvoluted peaks, colored by hierarchical density-based spatial clustering of applications with noise (HDBSCAN) cluster labels ($C_0$: blue, $C_1$: orange, $C_2$: green). The right-side panels present histograms of peak-center energies for each cluster, plotted as counts versus bias voltage (Supplementary Information 6). (d-g) 3D scatter plots of peak centers $c_{ijk}$ in (E, $\boldsymbol{r}_{ij}$) space: (d) all deconvoluted peaks ($C_0$–$C_2$), (e) $C_0$, (f) $C_1$, and (g) $C_2$. Peaks in $C_0$ are concentrated near-zero-energy and localized around vortex cores, while $C_1$ and $C_2$ exhibit broader distributions in space and energy.

Figure 3 presents the ML-assisted ZBP identification in FTS. After ruling out known and experimentally controllable extrinsic origins(Supplementary Information 1), the remaining non-splitting ZBPs observed at vortex cores are consistent with reported signatures of MZM[3]. The ML-assisted workflow can further filter out contributions from heterogeneity, such as subsurface defects, to achieve reliable ZBP detection and mitigate ZBP misidentification. Figure 3a shows the ZBC map $\rho(E = 0;\ \boldsymbol{r})$ acquired at 40 mK under $B_\perp = 2$ T. Figure 3b shows the deconvolution process for a representative dI/dV spectrum (orange marker in Fig. 3a) using multiple Lorentzian fittings. Apparent local peak positions in the LDOS curves are identified using a conventional second-derivatives method[28], used as initial parameters for multiple Lorentzian fitting[29]. Note that superconducting regions between vortices, which exhibit featureless subgap conductance, are excluded from the fitting[30]. We focus on complex in-gap states within $|V_{bias}| <$ 0.75 mV, an energy range compatible to the characteristic CdGM level spacing ($E_\mu \sim 0.74$ meV)[3, 17]. This energy window is chosen to focus on near zero-bias features that allow discrimination between ZBP signature and ZBP-mimicking states. The fit result for each spectrum is evaluated using the coefficient of determination ($R^2$), confirming subgap conductance near vortex cores is well described by multiple Lorentzian fittings (Supplementary Information 3).

After deconvolution, the extracted peak parameters are assembled into a structured feature set $\mathcal{F}$, with statistical outliers removed. The feature set is then processed through UMAP embedding followed by HDBSCAN clustering (Fig. 3c), and deconvoluted peaks are subsequently partitioned into three data-driven clusters: $C_0$, $C_1$, and $C_2$ (Supplementary Information 5). The right-side panels in Fig. 3c summarize the energy distributions of peak centers for each cluster, showing that $C_0$ is sharply concentrated near-zero-energy, while $C_1$ and $C_2$ exhibit broader and off-centered from zero bias (Supplementary Information 6). Figures 3d-g further visualize cluster-specific characteristics with 3D scatter plots of peak centers $c_{ijk}$ in (E, $\mathbf{r}_{ij}$) space. Peaks in $C_0$ are energetically concentrated near zero bias and localized around vortex cores (Fig. 3e), whereas $C_1$ and $C_2$ are more broadly distributed across energy and space (Fig. 3f, g). This cluster separation demonstrates that the unsupervised clustering can robustly distinguish ZBPs from other subgap states, providing a data-driven classification of spectral features.

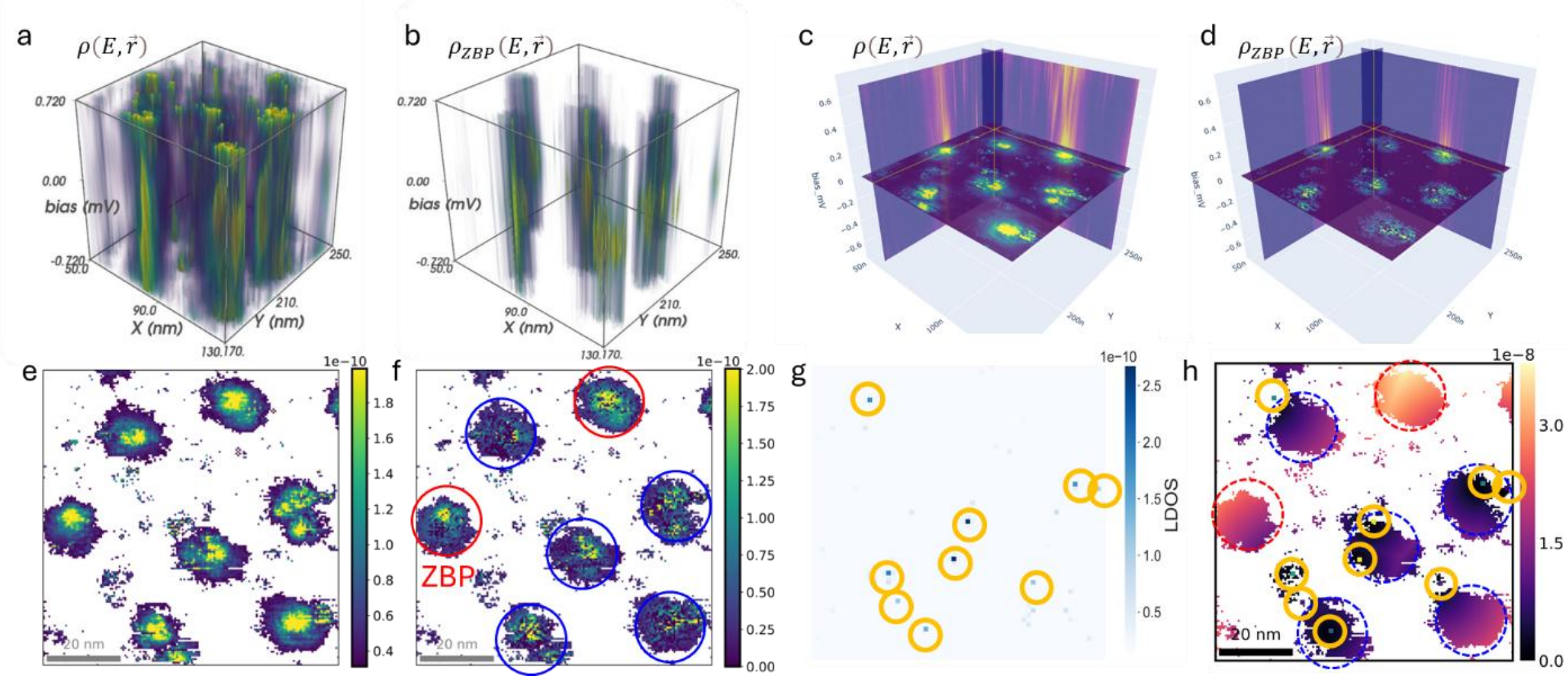

**Figure 4**. Reconstructed ZBC map and influence of subsurface defects. (a,b) 3D intensity map of (a) modeled LDOS $\rho(E,\boldsymbol{r})$ and (b) reconstructed ZBP-only LDOS $\rho_{ZBP}(E,\boldsymbol{r})$, visualizing subgap conductance near vortex cores under $B_\perp$ = 2 T. Bright regions indicate high LDOS, and superconducting regions are rendered transparent. The $\rho_{ZBP}(E,\boldsymbol{r})$ is reconstructed by summing the peak in ZBP cluster ($C_0$). (c,d) Orthogonal section views of (c) $\rho(E,\boldsymbol{r})$ and (d) $\rho_{ZBP}(E,\boldsymbol{r})$, the same datasets as (a,b). Vertical cross-sections correspond to x-E and y-E planes and slicing positions in (c) and (d) are the same. Horizontal planes show ZBC maps (c) $\rho(E=0,\boldsymbol{r})$ and (d) $\rho_{ZBP}(E=0,\boldsymbol{r})$. (e) ZBC map $\rho(E=0,\boldsymbol{r})$, $ZBC_{full}$, showing enhanced conductance at all vortex cores with transparent superconducting regions. (f) ZBC map $\rho_{ZBP}(E=0,\boldsymbol{r})$, $ZBC_{ZBP}$, discloses intensity variations among vortices: red circles mark robust and circular ZBP signatures; blue circles indicate suppressed or distorted ZBPs. (g) Zero-field ZBC map $\rho(E=0,\boldsymbol{r};\ B\ =\ 0\,T)$ acquired over the same field of view, highlighting local conductance enhancements associated with subsurface defects (yellow circled). (h) Distance-to-nearest-defect map derived from (g), showing spatial correlation between defect locations and the suppression of ZBPs.

After classifying spectral features based on peak properties, we examine how the resulting ZBP-consistent components are distributed in real space. By using ML-assigned cluster labels to filter the deconvoluted peaks, we isolate the ZBP contribution from complex in-gap states, as demonstrated in Fig. 2h. The ZBP-only LDOS $\rho_{ZBP}(E,\boldsymbol{r})$ is reconstructed by summing the Lorentzian components assigned to the ZBP-consistent cluster ($C_0$): $\rho_{ZBP}(E,\boldsymbol{r}) = \sum_{i,j,k\in C_0} p_k\left(E,\boldsymbol{r}_{ij}\right)$. Compared to the raw grid LDOS data ρ(E, $\mathbf{r}$) (Fig. 4a), the reconstructed $\rho_{ZBP}(E,\boldsymbol{r})$ (Fig. 4b) visualizes the spatial and energetic distribution of ZBP peaks and effectively removes non-ZBP components that complicate the discrimination of zero-bias features in the original data[31]. In the orthogonal sections (Figs. 4c,d), subgap conductance away from the vortex cores is strongly suppressed in $\rho_{ZBP}(E,\boldsymbol{r})$, leaving sharp zero-bias features localized at the vortex centers.

The ZBC map $\rho(E = 0, \boldsymbol{r})$ ($ZBC_{full}$, Fig. 4e) displays enhanced ZBC at all vortex cores, with the apparent ZBP-like signatures. In contrast, the $\rho_{ZBP}(E = 0, \boldsymbol{r})$ ($ZBC_{ZBP}$, Fig. 4f) reveals that only a subset of vortices (red circles) exhibit pronounced, circular-shaped ZBC enhancements, while others (blue circles) appear suppressed or distorted. This contrast demonstrates that the part of the apparent ZBC enhancement in the $ZBC_{full}$ comes from near-zero, off-centered excitations, which are not classified as $C_0$. These contributions from $C_1$ or $C_2$ are ruled out in the $ZBC_{ZBP}$, removing ZBP-mimicking features. Accordingly, the only vortices with red circles in Fig. 4f exhibit ZBP features that are consistent with established experimental signatures of MZMs, whereas the remaining vortices are attributed to trivial in-gap states (Supplementary Information 7).

To understand the contrast between $ZBC_{full}$ and $ZBC_{ZBP}$, we further examine local heterogeneities that could shift CdGM states toward zero energy [17]. A zero-field LDOS map $\rho(E = 0, \boldsymbol{r}\,; B = 0\ T)$ (Fig. 4g) is used to identify nearly zero-energy states associated with disorder, such as subsurface defects without magnetic field. We register the defect positions in yellow circles in Fig. 4g, which is obtained on the same field of view after drift correction algorithm (Supplementary Information 8). Figure 4h presents the distance-to-nearest-defect map overlaid with indicators of MZM-candidate vortices (red circles) from $ZBC_{ZBP}$ and defect locations (yellow circles) from Fig. 4g, respectively. Interestingly, the vortices with reduced ZBPs tend to be located closer to the defects. This spatial proximity, highlighted by blue circles in $ZBC_{ZBP}$, suggests an association between local heterogeneity and the suppression or distortion of vortex-core ZBPs. This correlation confirms that disorder-induced shifts of CdGM states can modify the apparent ZBP features and may contribute to the vortex-to-vortex variability in ZBP robustness reported across different sites.

## Conclusion

We have developed a data-driven, ML-assisted workflow to decipher MZM signatures in TSC using high spatial and energy resolution grid LDOS data. We deconvolute the complex subgap conductance, assemble peak parameters into a feature set, and use unsupervised ML clustering to precisely identify ZBP-consistent signals in FTS. We detect ZBP features through classification of grid LDOS data based on spectral properties and distinguish them from ZBP-mimicking near-zero-energy states. The reconstructed ZBP-only grid LDOS data enables systematic identification of MZM-candidate ZBPs while mitigating ZBP misidentification arising from trivial near-zero-energy states. This analysis further reveals the impacts of local heterogeneities on ZBP distribution. Moreover, ML-assisted workflow can be extended by applying deep-learning architectures[32] or by classifying complex in-gap states to examine their spatial distributions across diverse topological heterostructure systems. The data-driven approach establishes a reproducible and scalable framework applicable to diverse quantum materials and experimental conditions. Accordingly, the workflow can be extended to complementary probes, including nonlocal transport and spin-polarized STM, to strengthen MZM identification and to track the evolution of ZBP distributions under various conditions with precise spectral and spatial separation of complex in-gap states. This workflow provides a robust foundation for the systematic mapping and manipulation of MZMs in TSC, thereby providing a practical route to more reliable STM-based identification of MZMs relevant to topological quantum computing efforts.

# Methods

## STM/S measurements

We utilized a dilution refrigerator STM (UNISOKU USM1600) operating at 40 mK equipped with vector field (2-2-9 T) magnet. The $FeTe_{0.55}Se_{0.45}$ single crystal[24] was cleaved at 83 K under ultra-high vacuum condition ($< 1\times10^{-10}$ Torr) and immediately transferred to a pre-cooled STM head (4.2 K). The sample temperature reached 40 mK after further cooling with dilution refrigerator. We used a mechanically polished PtIr tip or an electrochemically etched W tip, and electronic states of the tips were verified on an Au(111) single crystal.

## Data analysis

The STM data were analyzed using Python scripts incorporating ML and image processing packages. During this work, the authors used the ORNL AI-assist program to support data-analysis code development and its documentation. All outputs were reviewed and verified by the authors, who take full responsibility for the final content.

## Data availability

The data that support the findings of this study are available from the corresponding author upon request.

## Code availability

Python scripts used for the analysis are available in Github([https://github.com/jewook-park/ZBPs_in_FTS](https://github.com/jewook-park/ZBPs_in_FTS))

# Acknowledgement

The research was supported by the Center for Nanophase Materials Sciences, (CNMS), which is a US Department of Energy, Office of Science User Facility at Oak Ridge National Laboratory. Crystal growth and characterization was supported by the U.S. Department of Energy, Office of Science, Basic Energy Sciences, Materials Sciences and Engineering Division.